\documentclass[a4paper,fleqn]{cas-dc}

\usepackage[numbers]{natbib}
\usepackage[utf8]{inputenc}
\DeclareUnicodeCharacter{22EF}{\dots}

\usepackage{soul}
\def\tsc#1{\csdef{#1}{\textsc{\lowercase{#1}}\xspace}}
\tsc{WGM}
\tsc{QE}
\tsc{EP}
\tsc{PMS}
\tsc{BEC}
\tsc{DE}

\begin{document}
\let\WriteBookmarks\relax
\def\floatpagepagefraction{1}
\def\textpagefraction{.001}
\shorttitle{Graphenyldiene-H$_2$}
\shortauthors{Laranjeira et~al.}

\title [mode = title]{Potassium Decoration on Graphenyldiene Monolayer for Advanced Reversible Hydrogen Storage}

\author[1]{José A. S. Laranjeira}
\affiliation[1]{
organization={Modeling and Molecular Simulation Group},
addressline={São Paulo State University (UNESP), School of Sciences}, 
city={Bauru},
postcode={17033-360}, 
state={SP},
country={Brazil}}
\credit{Conceptualization of this study, Methodology, Review and editing, Investigation, Formal analysis, Writing -- review \& editing, Writing -- original draft}

\author[1]{Nicolas F. Martins}
\credit{Data curation, Formal analysis, Writing -- review \& editing, Writing -- original draft}

\author[2]{Kleuton A. L. Lima}
\affiliation[2]{
organization={Department of Applied Physics and Center for Computational Engineering and Sciences},
addressline={State University of Campinas}, 
city={Campinas},
postcode={13083-859}, 
state={SP},
country={Brazil}}
\credit{Conceptualization of this study, Methodology, Review and editing, Investigation, 
Formal analysis, Writing -- review \& editing, Writing -- original draft}

\author[3,4]{Bill. D. Aparicio-Huacarpuma}
\affiliation[4]{
organization={Institute of Physics},
addressline={University of Brasília}, 
city={Brasília },
postcode={70910‑900}, 
state={DF},
country={Brazil}}
\affiliation[5]{
organization={Computational Materials Laboratory, LCCMat, Institute of Physics},
addressline={University of Brasília}, 
city={Brasília },
postcode={70910‑900}, 
state={DF},
country={Brazil}}
\credit{Conceptualization of this study, Methodology, Review and editing, Investigation, 
Formal analysis, Writing -- review \& editing, Writing -- original draft}

\author[3,4]{Luiz A. Ribeiro Junior}
\credit{Conceptualization of this study, Methodology, Review and editing, Investigation, 
Formal analysis, Writing -- review \& editing, Writing -- original draft}

\author[5]{Xihao Chen}
\affiliation[5]{
organization={School of Materials Science and Engineering},
addressline={Chongqing University of Arts and Sciences}, 
city={Chongqing},
postcode={402160}, 
country={China}}
\credit{Conceptualization of this study, Methodology, Review and editing, Investigation, 
Formal analysis, Writing -- review \& editing, Writing -- original draft}

\author[2]{Douglas S. Galvao}
\credit{Conceptualization of this study, Methodology, Review and editing, Investigation, Formal analysis, Writing -- review \& editing, Writing -- original draft}

\author[1]{Julio R. Sambrano}
\ead{jr.sambrano@unesp.br}
\cormark[1]
\cortext[cor1]{Corresponding author}
\credit{Conceptualization of this study, Methodology, Review and editing, Investigation, 
Formal analysis, Writing -- review \& editing, Writing -- original draft}

\begin{abstract}
Potassium-decorated graphenyldiene (K@GPD) is investigated as a promising two-dimensional material for reversible hydrogen storage using first-principles density functional theory calculations. Potassium atoms bind strongly to the GPD monolayer, and ab initio molecular dynamics (AIMD) simulations confirm the thermal stability of the functionalized system at 300 K. Hydrogen adsorption energies range from -0.11 to -0.14 eV per H$_2$, denoting reversible storage. At full coverage (18 H$_2$ molecules), the system reaches a storage capacity of 8.82 wt\%, exceeding the U.S. DOE target. AIMD simulations reveal spontaneous H$_2$ desorption at ambient temperature, demonstrating excellent reversibility. 
\end{abstract}


\begin{highlights}
\item K-decorated graphenyldiene enables reversible H$2$ storage with optimal adsorption energies
\item Hydrogen storage capacity of 8.82 wt\% surpasses DOE target
\item AIMD simulations confirm thermal stability and reversible H$_2$ desorption at 300 K
\item K@GPD is a promising 2D platform for clean and efficient hydrogen energy systems
\end{highlights}


\maketitle

\section{Introduction}

The environmental impact of greenhouse gas emissions, particularly carbon dioxide released during the extraction, processing, and combustion of fossil fuels, has prompted intense research into advanced materials and coatings for sustainable energy storage solutions \cite{Kim2020Novel, Dalwadi2023The, Ramasubramanian2022Towards, Chen2024MXene}. Hydrogen, one of the most abundant elements in the universe, offers a high energy density, making it a promising carbon-neutral energy carrier with strong potential to address global environmental challenges \cite{Evro2024Carbon, Kovač2021Hydrogen, Evangelopoulou2019Energy, Highfield2025Sustainable}.

Achieving net-zero carbon dioxide emissions by 2050 through decarbonizing industrial activities has become a central international objective \cite{Lau2023Global, Rissman2020Technologies, Schreyer2020Common}. To facilitate this transition, the U.S. Department of Energy (DOE) has set a target for onboard hydrogen storage systems to reach a gravimetric capacity of at least 5.5 wt\% under ambient conditions by 2025 \cite{HUZAIFA2025114915}.

Hydrogen presents several advantages as a clean energy carrier, but storage challenges hinder its widespread adoption \cite{REHMAN2025112497}. A range of storage strategies has been investigated, including liquefaction \cite{luo2024study}, high-pressure gas cylinders \cite{li2024comparative}, and solid-state approaches \cite{xu2024research}. Nevertheless, both liquefied and compressed hydrogen storage methods are associated with high costs and limited long-term efficiency \cite{Bian2021Thermodynamic, Jouybari2022Thermodynamic}. Maintaining hydrogen in liquid form demands extremely low temperatures and high pressures, which introduces substantial technical and economic burdens. Moreover, due to its small molecular size, hydrogen poses a high risk of leakage from storage systems \cite{Du2024Leakage}.

Solid-state and electrochemical methods have emerged as viable alternatives for hydrogen storage \cite{tang2023state}. Metal hydrides and physisorption-based systems, particularly those employing materials with high surface areas, have demonstrated encouraging results, although further optimization is still needed \cite{LUO2020100071, BORETTI2025100226}. Despite substantial progress, identifying ideal storage materials remains a challenging task \cite{ghotia2025review}. Recent advances suggest that strategies such as elemental doping \cite{cid2022enhanced, kundu2022yttrium, tang2024dft}, tuning reaction pathways \cite{johnson2022exploring, ali2021modification}, and controlling particle size through advanced fabrication techniques \cite{HA20251047} may significantly enhance storage performance.

Two-dimensional (2D) materials exhibit unique physicochemical properties that make them attractive for applications in energy storage, electronics, and catalysis~\cite{fan2021emerging, hayat2023recent, xie2021chemistry, qin2023two}. However, their chemically inert surfaces limit their intrinsic capacity for hydrogen adsorption. To address this limitation, surface functionalization with metal atoms has been extensively investigated. Alkali metals~\cite{habibi2021reversible, XU2024226, NABI2024117742, SOSA202120245}, alkaline earth metals~\cite{KHAN2023107471, ZHANG2024136, MA2021101985, ARELLANO202120266}, and transition metals~\cite{app11146604, en17163944, DARVISHNEJAD202440, BOEZAR202138370} have been shown to enhance hydrogen binding energies by introducing active adsorption sites. These dopants facilitate interactions between hydrogen molecules and the 2D substrate, improving storage capacity while maintaining moderate binding strength.

Motivated by the growing interest in carbon-based 2D frameworks and their application in green energy solutions \cite{titirici2015sustainable}, we recently proposed a novel planar $sp^2$-hybridized carbon monolayer referred to as graphenyldiene (GPD) using density functional theory (DFT) calculations \cite{LARANJEIRA2024100321}. This structure features a distinctive topology of 18-, 6-, and 4-membered carbon rings arranged in a periodic hexagonal lattice. Its backbone incorporates Dewar-benzene motifs instead of the cyclobutadiene units present in graphenylene. This results in a porous network with uniformly distributed octadecagonal pores, each with a diameter of approximately 8.32~\AA.

The calculations have also indicated that GPD possesses a direct band gap of 1.26~eV, making it suitable for semiconducting applications. The monolayer also exhibits high carrier mobility, on the order of $10^3$~cm$^2$/V$\cdot$s, suggesting potential for use in nanoelectronic devices. Molecular dynamics simulations confirmed its thermal stability, with the structure remaining intact up to 1000~K. Additionally, GPD has been proposed as a potential precursor to graphdiyne, undergoing a predicted phase transition near 1500~K, underscoring it.

Building on these insights, this work explores the potential of potassium-decorated GPD (K@GPD) as an efficient platform at ambient temperature for reversible hydrogen storage through DFT calculations. The study evaluates key physicochemical properties, including adsorption energetics, thermal stability, charge transfer, and electronic structure. The results demonstrate that K@GPD exhibits suitable characteristics for physisorption-based hydrogen storage, with performance metrics that place it as a promising 2D candidate.

\section{Methodology}

First-principles DFT-based calculations were carried out to investigate the structural and electronic properties of GPD and evaluate its suitability for hydrogen storage. The generalized gradient approximation (GGA) with the Perdew–Burke–Ernzerhof (PBE) functional~\cite{PhysRevLett.77.3865, ernzerhof1999assessment} and the projector augmented wave (PAW) method~\cite{PhysRevB.50.17953} were employed, as implemented in the Vienna \textit{ab initio} Simulation Package (VASP). Electronic wave functions were expanded in a plane-wave basis set with a kinetic energy cutoff of 520~eV. A vacuum spacing of 15~\r{A} was applied along the $z$-direction to eliminate spurious interactions between periodic images.

Structural optimizations and projected density of states (PDOS) calculations used $\Gamma$-centered \textbf{k}-point meshes with a $3 \times 3 \times 1$ grid size. Dispersion interactions were accounted for using the DFT-D2 correction scheme proposed by Grimme~\cite{grimme2006semiempirical}. Structural relaxation employed the conjugate gradient algorithm, with convergence criteria set to a total energy variation below $ 1\times10^ {-5}$~eV and Hellmann–Feynman forces smaller than 0.01~eV/\r{A} for all atoms. Charge transfer was analyzed using the Bader decomposition method. Ab initio molecular dynamics (AIMD) simulations were performed at 300~K for 5~ps with a time step of 0.5~fs, controlled by the Nosé thermostat~\cite{hoover1985canonical}.

The charge density difference (CDD) obtained from the K@GPD system was calculated by:

\begin{equation}
\Delta \rho = \rho_{\text{(K@GPD})} - \rho_{\text{(K)}} - \rho_{\text{(K@GPD)}},
\label{eq:charge_density}
\end{equation}

\noindent where $\rho_{(\text{K@GPD})}$, $\rho_{(\text{K})}$, and $\rho_{K@GPD}$ refer to the charge densities of the K@GPD substrate, isolated K adatoms, and pristine GPD monolayer, respectively. For the hydrogen-adsorbed K@GPD system, the CDD was obtained as:

\begin{equation}
\Delta \rho = \rho_{(\text{K@GPD + H$_2$})} - \rho_{(\text{H$_2$})} - \rho_{(\text{K@GPD})},
\label{eq:charge_density1}
\end{equation}

\noindent where $\rho_{(\text{K@GPD + H$_2$})}$, $\rho_{(\text{H$_2$})}$, and $\rho_{(\text{K@GPD})}$ refer to the charge densities of the K@GPD with H$_2$ molecules, isolated H$_2$ molecules, and K@GPD substrate, respectively.

The adsorption energy ($E_{\text{ads}}$) for the K@GPD + $n \textrm{H}_2$ systems was calculated using the following expression:

\begin{equation}
\begin{split}
E_{\text{ads}} &= \\\frac{1}{n} &\Big( E_{\text{K@GPD + $n \textrm{H}_2$}}
\quad - E_{\text{K@GPD}} - n \textrm{H}_2 \Big),
\end{split}
\end{equation}

\noindent where $E_{\text{K@GPD + $n \textrm{H}_2$}}$ represents the total energy of the K@GPD system with $n$ adsorbed H$_2$ molecules, $E_{\text{K@GPD}}$ is the energy of the bare K@GPD substrate, and $E_{\text{H}_2}$ corresponds to the energy of an isolated H$_2$ molecule.

The hydrogen adsorption capacity (HAC) in weight percentage was calculated as:

\begin{equation}
\text{HAC} (\text{wt\%}) = \frac{ n_\text{H} M_\text{H}}{n_\text{C} M_\text{C} + 
n_\text{K} M_\text{K} + n_\text{H} M_\text{H} },
\end{equation}

\noindent where $n_X$ and $M_X$ represent the number of atoms and molar masses of element $X$ ($X = \text{H}, \text{C}, \text{K}$), respectively.

Assuming atmospheric pressure (1 atm), the hydrogen desorption temperature ($T_{\text{R}}$) was estimated using the van't Hoff equation \cite{durbin2013review, alhameedi2019metal}:

\begin{equation}
T_{\text{des}} = \left| E_{\text{ads}} \right|\frac{R}{K_B \Delta S},
\end{equation}

\noindent where $R$ is the universal gas constant, $k_B$ is the Boltzmann constant, and $\Delta S$ represents the entropy change associated with the phase transition of hydrogen from the gas phase to the liquid phase (75.44 J mol$^{-1}$ K$^{-1}$).

A thermodynamic analysis was performed to evaluate the adsorption and desorption behavior of H$_2$ molecules under realistic conditions, employing the grand canonical partition function $Z$, given by:

\begin{equation}
   Z = 1 + \sum_{i=1}^{n} \exp\left(-\frac{E_i^{\text{ads}} - \mu}{k_B T}\right).
\end{equation}

In this equation, $n$ denotes the total number of H$_2$ molecules that can be adsorbed, while $\mu$ represents the chemical potential of a hydrogen molecule in the gaseous state. Here, $E_{i}^{\text{ads}}$ corresponds to the adsorption energy of the $i$-th H$_2$ molecule \cite{hashmi2017ultra, kaewmaraya2023ultrahigh}.

\section{Results and Discussion}

\begin{figure*}[pos=htb!]
    \centering
    \includegraphics[width=1\linewidth]{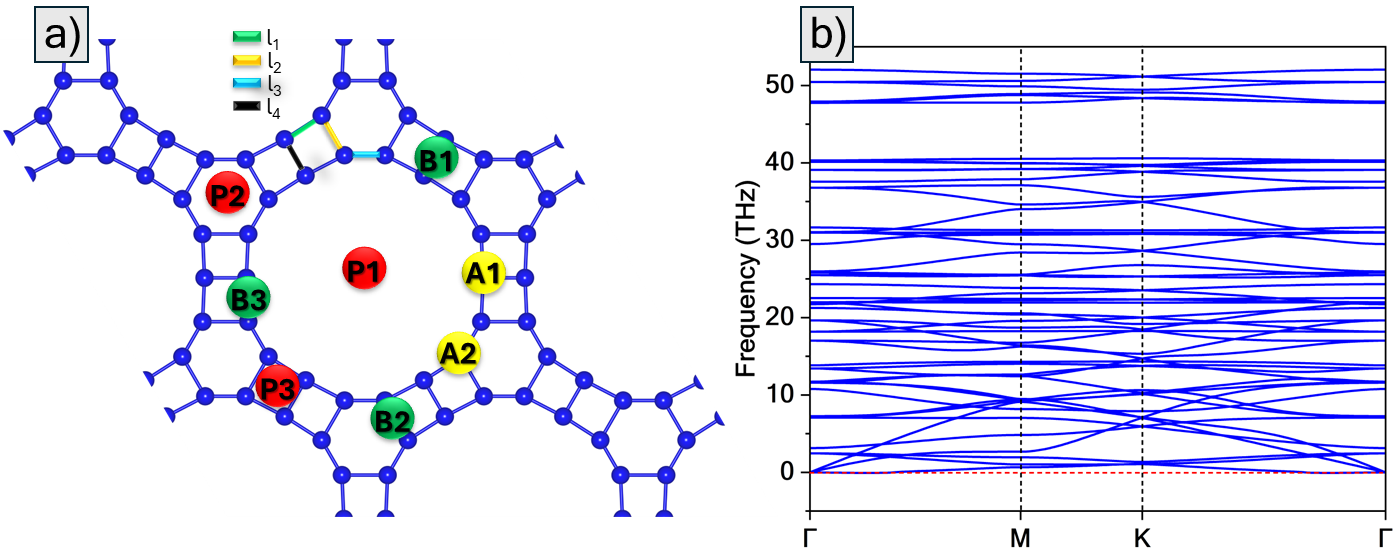}
    \caption{(a) Top view of the GPD monolayer highlighting the hexagonal 2 $\times$ 2 supercell and the evaluated high-symmetry adsorption sites for K decoration. Sites labeled A1--A2 indicate atomic sites, B1--B3 correspond to bond positions, while P1--P3 are related to pore sites. The optimized structure exhibits four distinct bond lengths: $l_1 = 1.47$~\AA (green), $l_2 = 1.53$~\AA (yellow), $l_3 = 1.35$~\AA (blue), and $l_4 = 1.40$~\AA (black). (b) Phonon dispersion of pristine GPD along high-symmetry paths in the Brillouin zone.}
    \label{fig:cell+phonon}
\end{figure*}

GPD crystallizes in a hexagonal unit cell with space group $\mathrm{P6/mmm}$ (No.~191) and lattice parameters $a = b = 9.32$~\AA. At the DFT/PBE level, the calculated cohesive energy is $-6.92$~eV/atom, in close agreement with the value reported in reference \cite{LARANJEIRA2024100321} ($-7.30$~eV/atom). The optimized structure exhibits four distinct bond lengths: $l_1 = 1.47$~\AA~(green), $l_2 = 1.53$~\AA~(yellow), $l_3 = 1.35$~\AA~(blue), and $l_4 = 1.40$~\AA~(black). The 18-membered rings show a diameter of approximately 8.33~\AA, forming a regular porous network, as illustrated in Fig.~\ref{fig:cell+phonon}a. This figure depicts a $2 \times 2$ supercell, where the adsorption sites considered for potassium decoration are indicated. These include three pore-centered sites (P1--P3), three bridge sites between hexagons and tetragons (B1--B3), and two top sites above carbon atoms (A1 and A2). 

Fig.~\ref{fig:cell+phonon}b shows the phonon dispersion of GPD along the high-symmetry path of the Brillouin zone. The absence of imaginary modes confirms the dynamical stability of the monolayer within the computational framework adopted. Distinct phonon branches and a phononic band gap in the range of 40--47~THz are observed.

\begin{figure*}[pos=htb!]
    \centering
    \includegraphics[width=0.8\linewidth]{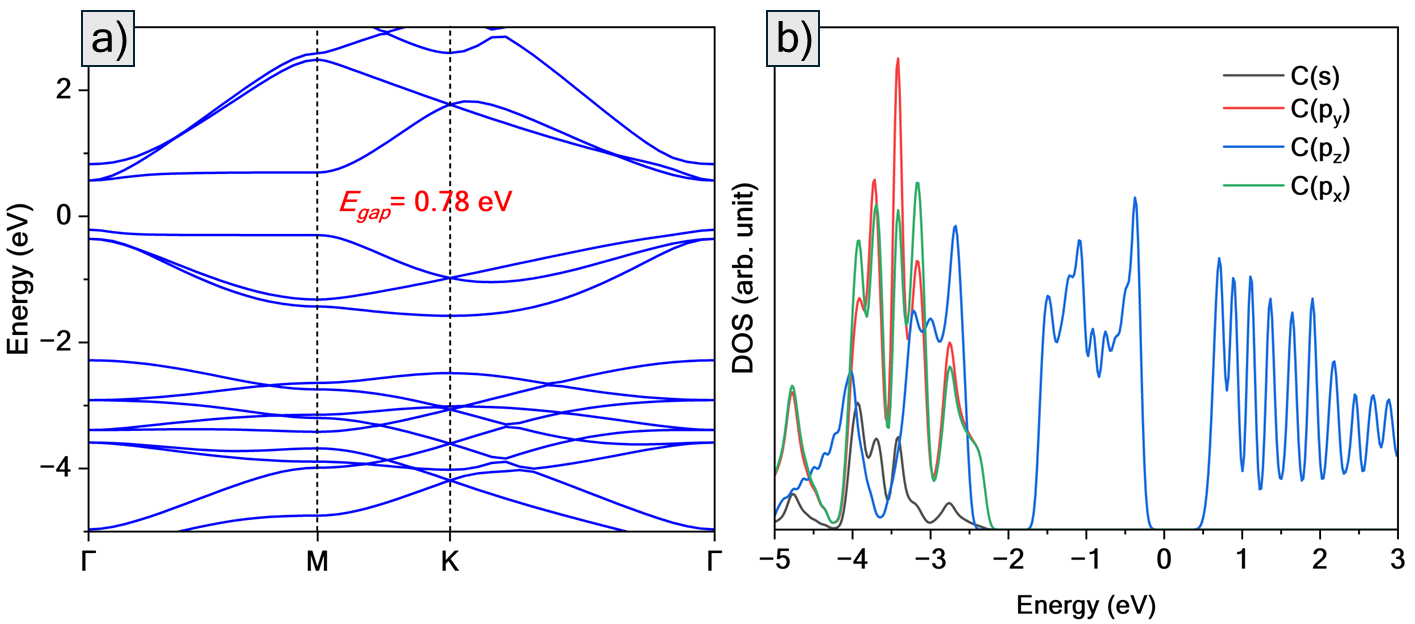}
    \caption{(a) Band structure and (b) PDOS for GPD system. The monolayer exhibits semiconducting behavior, characterized by a direct band gap transition ($\Gamma \rightarrow \Gamma$) of 0.78 eV calculated at DFT/PBE level.}
    \label{fig:band+dos}
\end{figure*}

The electronic band structure and projected density of states (PDOS) of GPD are presented in Fig.~\ref{fig:band+dos}. This 2D carbon allotrope displays a direct band gap of 0.78~eV at the $\Gamma$ point. This value is lower than the 1.26~eV reported in reference \cite{LARANJEIRA2024100321}, which used a double-zeta valence with polarization (DZVP) gaussian basis set combined with the HSE06 hybrid functional. The discrepancy arises from the known tendency of the PBE functional to underestimate band gaps. Nevertheless, despite methodological differences, the overall band dispersion obtained here strongly agrees with previous results.

The band structure reveals nearly flat regions at the valence band maximum (VBM) and conduction band minimum (CBM) along the $\Gamma \rightarrow M$ path and two cone-like crossings at the $K$ point. These features indicate that tuning the chemical potential could substantially influence carrier mobility in the GPD monolayer.

The PDOS analysis reveals that the deeper valence bands receive contributions from all carbon orbitals, with C($p_x$) and C($p_y$) dominating the $\sigma$-bonding states. Following a gap in the density of states between approximately $-2.2$~eV and $-1.8$~eV, the states near the VBM are primarily derived from C($p_z$) orbitals. The abrupt decrease in PDOS at the VBM reflects the flat dispersion observed in this energy range. As previously reported~\cite{LARANJEIRA2024100321}, these states correspond to anti-bonding $\pi$ orbitals localized along the intra-pore bonds of the hexagonal rings. In the conduction band, C($p_z$) orbitals dominate, now associated with bonding $\pi$ states mainly distributed along the connections between hexagonal and four-membered rings.

Potassium adsorption on the GPD monolayer was evaluated by initially placing K atoms on the eight non-equivalent adsorption sites illustrated in Fig.~\ref{fig:cell+phonon}. These sites are classified as atom sites (A1 and A2), bond sites (B1, B2, and B3), and pore sites (P1, P2, and P3). The corresponding adsorption energies and relaxed configurations are summarized in Table~\ref{tab:sites}. The lowest adsorption energy, $-2.62$~eV, corresponds to the P2 site, followed closely by P1 ($-2.61$~eV); both configurations are nearly degenerate and energetically favorable. Interestingly, initial placements at A1, B1, B3, and P3 all relaxed to the B1 site, yielding a slightly less stable configuration with an energy of $-2.58$~eV. All adsorption energies are more negative than the cohesive energy of bulk potassium ($-0.94$~eV/atom)~\cite{averill1972calculation, JIANG2024865}, indicating that isolated K adatoms preferentially bind to GPD rather than forming metallic clusters. Based on these results, the P2 site was selected for potassium decoration, forming the K@GPD system with four K atoms per unit cell.

\begin{table}[!ht]
\centering
\caption{Adsorption energies (E$_\text{ads}$) and final configurations for the adsorption sites evaluated during K decoration on the GPD nanosheet.}
\label{tbl2}
\begin{tabular*}{\linewidth}{@{\extracolsep{\fill}} lcc}
\toprule
\textbf{Initial site} & \textbf{E$_\text{ads}$ (eV)} & \textbf{Final site} \\
\midrule
A1 &  -2.58  & B1  \\
A2 &  -2.62  & P2  \\
B1 &  -2.58  & B1  \\
B2 &  -2.62  & P2   \\
B3 &  -2.58  & B1  \\
P1 &  -2.61  & P1  \\
P2 &  -2.62  & P2  \\
P3 &  -2.58 & B1  \\
\bottomrule
\end{tabular*}
\label{tab:sites}
\end{table}

\begin{figure*}[pos=htb!]
    \centering
    \includegraphics[width=0.8\linewidth]{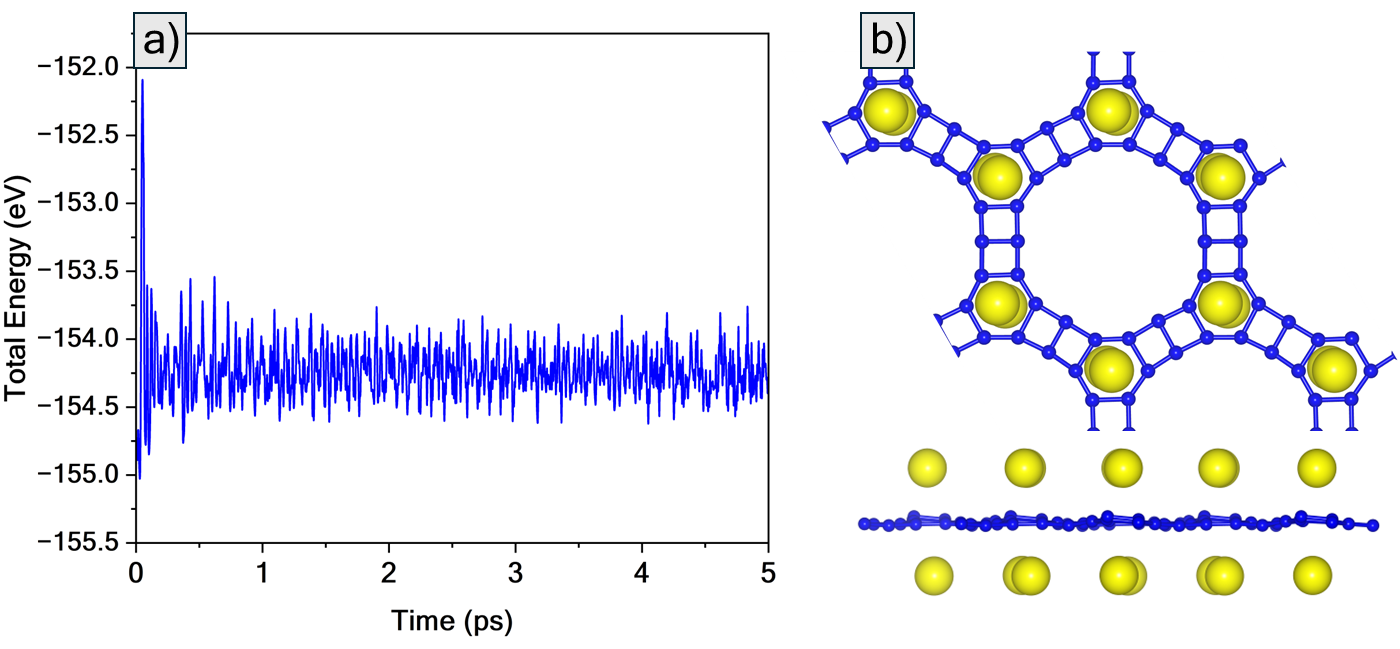}
    \caption{AIMD simulations results for 300 K by 5 ps in K@GPD system. (a) Energy fluctuations over 5 ps during the simulation. (b) Final structure obtained at the end of the simulations.}
    \label{fig:aimd_decorated}
\end{figure*}

The thermal stability of the K@GPD system was assessed via AIMD simulations at 300~K for 5~ps, as shown in Fig.~\ref{fig:aimd_decorated}. The total energy fluctuates around a mean value of approximately $-154.25$~eV, with no indications of structural rearrangements, potassium desorption, or phase transitions. The inset confirms that K adatoms remain bound to their most favorable adsorption sites, exhibiting only minor thermal fluctuations. Slight out-of-plane displacements of carbon atoms are also observed, attributed to vibrational motion at finite temperature. These findings demonstrate the thermal robustness of the K@GPD system under ambient conditions.

This behavior aligns with previous reports on K-decorated carbon-based materials. In K-decorated DHP-graphene~\cite{WANG2025141932}, AIMD simulations similarly confirmed stable potassium retention at room temperature, supporting a hydrogen storage capacity of 8.1~wt\%. K-decorated C$_9$N$_4$ also exhibited strong K adsorption and promising storage characteristics~\cite{KAUR202326301}.

\begin{figure*}[pos=htb!]
    \centering
    \includegraphics[width=0.8\linewidth]{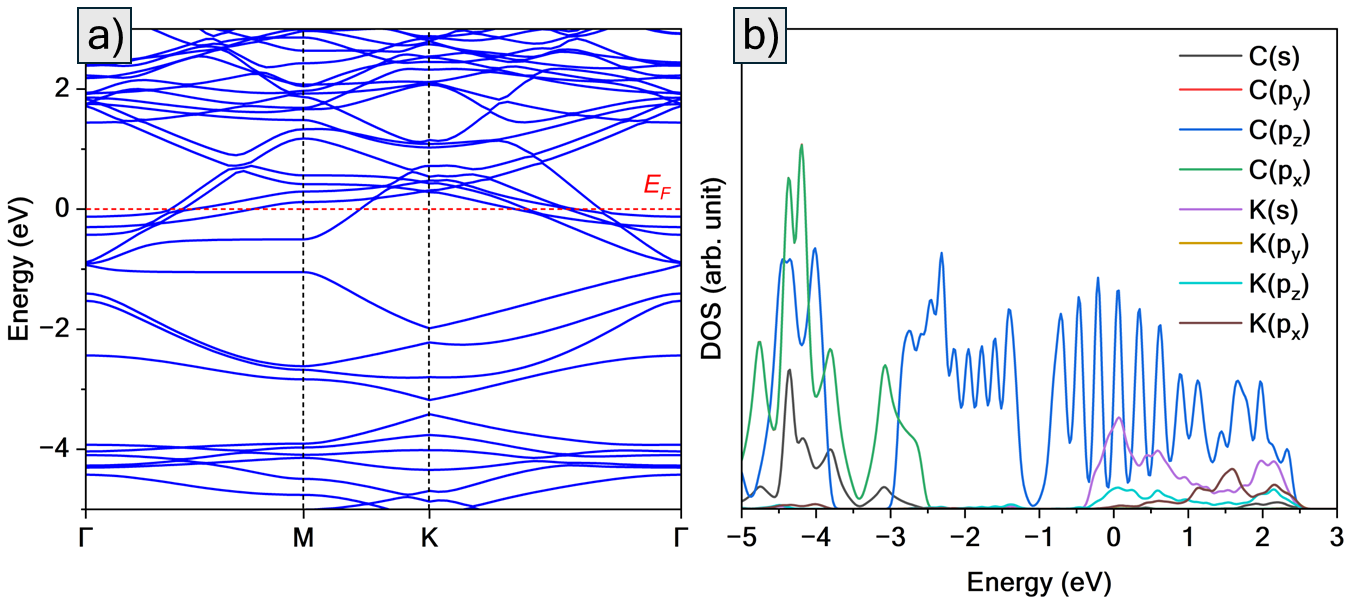}
    \caption{(a) Band structure and (b) PDOS for K@GPD system. The system exhibits metallic behavior, characterized by several bands that cross the Fermi level ($E_F$ red dashed line).}
    \label{fig:band_dos_decorated}
\end{figure*}

The electronic band structure and related PDOS of K@GPD are presented in Fig.~\ref{fig:band_dos_decorated} to analyze the effects of potassium adsorption on the electronic properties of GPD. Upon decoration, the material transitions from a semiconductor to a metallic state, as indicated by the emergence of several bands crossing the Fermi level ($E_F$, shown as a red dashed line). This metallic behavior reflects the partially occupied electronic states introduced by potassium atoms.

It is worth mentioning that in the PDOS analysis, at lower energies within the valence band, the electronic structure remains essentially unchanged compared to the pristine monolayer, with prominent contributions from $\sigma$ states associated with C($s$) and C($p_x$) orbitals. Above approximately $-2.5$~eV, the states are predominantly derived from C($p_z$) orbitals, with minimal input from other atomic components. Near the Fermi level, new partially occupied states appear, primarily involving K($s$), K($p_x$), K($p_z$), and C($p_z$) orbitals, indicating strong hybridization between the potassium adatoms and the carbon $\pi$ system. This hybridized character extends into the conduction band, where these contributions continue to dominate.

\begin{figure}[pos=htb!]
    \centering
    \includegraphics[width=0.8\linewidth]{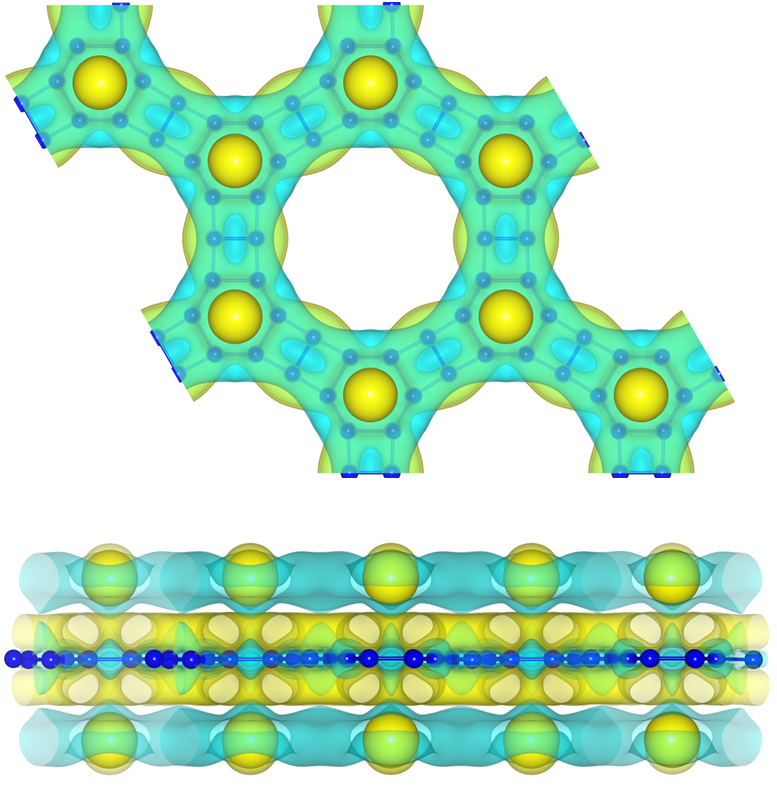}
    \caption{Top and side views of the Charge density difference map for K@GPD complex. The yellow (blue) regions indicate charge accumulation (depletion).}
    \label{fig:cdd}
\end{figure}

The interaction between potassium adatoms and the GPD monolayer was further examined by calculating the charge density difference (CDD), as shown in Fig.~\ref{fig:cdd}. This map's yellow and blue regions denote charge accumulation and depletion. A clear depletion zone appears directly above each K atom. At the same time, pronounced accumulation regions are located above and below the carbon plane, reflecting the redistribution of electronic density induced by adsorption.

Within the monolayer, localized charge depletion is also observed, particularly on the $\pi$ orbitals of the carbon atoms. This pattern indicates that the charge transferred from potassium is primarily delocalized over the conjugated $\pi$ system, suggesting strong electrostatic coupling mediated by orbital overlap between the adatoms and the carbon network.

Quantitative confirmation of this mechanism is provided by Bader charge analysis, which shows that each K atom donates approximately -0.50~$|e|$ to the GPD surface. This electron transfer behavior is consistent with previous studies on alkali-metal-decorated carbon materials, where similar charge donation has been associated with the formation of active adsorption sites for hydrogen storage~\cite{NABI2024117742, WANG2025141932, LARANJEIRA2025139, LIU2025105802}.

\begin{figure*}[pos=htb!]
    \centering
    \includegraphics[width=0.8\linewidth]{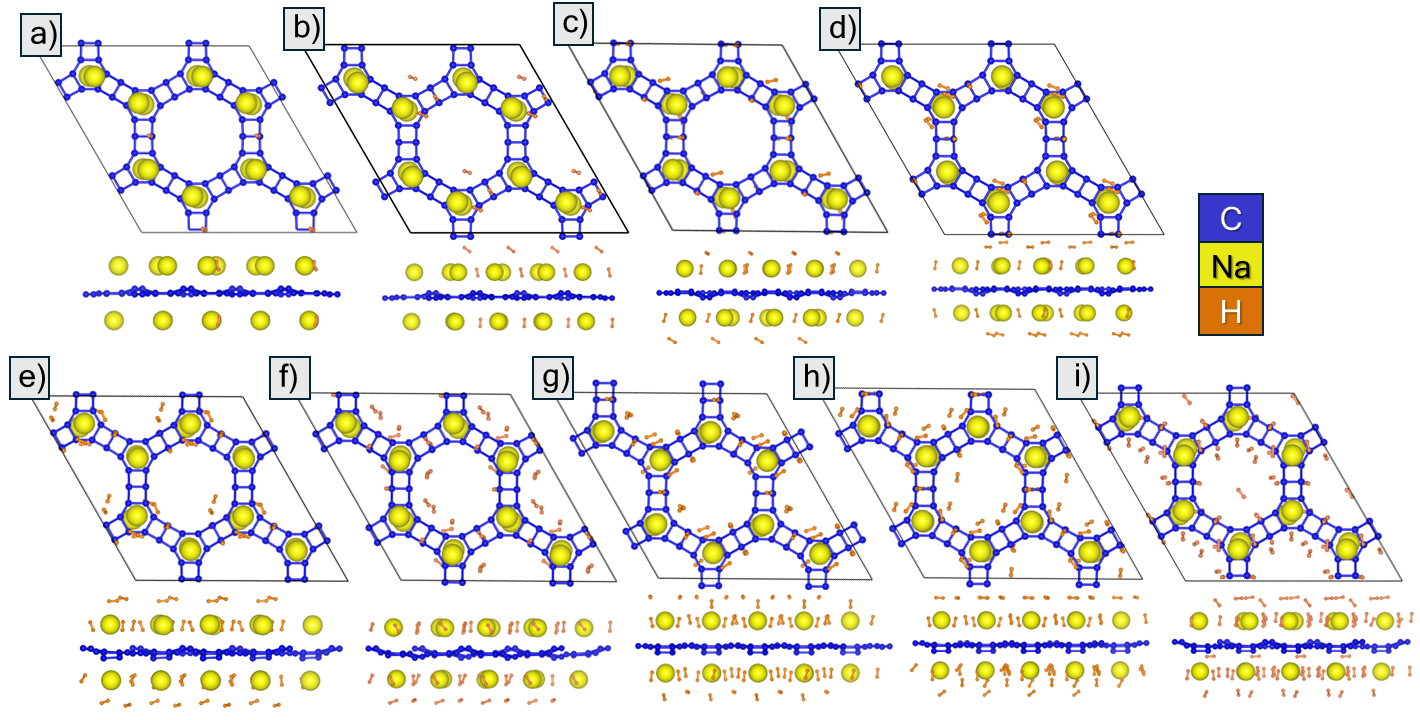}
    \caption{H$_2$ saturation on K@GPD pathway, where (a), (b), (c), (d), (e), (f), (g), (h), and (i) denote K@GPD + 2H$_2$, 4H$_2$, 6H$_2$, 8H$_2$, 10H$_2$, 12H$_2$, 14H$_2$, 16H$_2$, and 18H$_2$ molecules, respectively.}
    \label{fig:h2_saturation}
\end{figure*}

The hydrogenation process of the K@GPD system was systematically investigated by introducing H$_2$ molecules in pairs, ranging from 2H$_2$ to 18H$_2$, as illustrated in Fig.~\ref{fig:h2_saturation}. With increasing H$_2$ coverage, the GPD nanosheet exhibited progressive structural distortion, characterized by pronounced buckling, while the K adatoms retained their stable adsorption sites, unaffected by hydrogenation.  

Critical parameters of the process, including adsorption energy ($E_\mathrm{ads}$), average and maximum H–H bond length ($R_\mathrm{H-H}$ and R$_\mathrm{H-H_{Max}}$), desorption temperature ($T_D$), and hydrogen storage capacity (HSC), are summarized in Table~\ref{tab:h2_adsorption}. The $E_\mathrm{ads}$ values remained nearly constant ($-0.11$ to $-0.14$ eV), indicating non-cooperative H$_2$ adsorption, where intermolecular interactions play a negligible role. An irregular $E_\mathrm{ads}$ behavior regarding H$_2$ concentration is also observed by Kaur \textit{ et al.} \cite{KAUR202326301} in K@C$_9$N$_4$ complex. This probably stems from the large atomic radius of K, which optimizes H$_2$ distribution around active sites. These $E_\mathrm{ads}$ values are within the reversible hydrogen storage range ($-0.5$ to $-0.1$ eV) \cite{LIU2025105802, ABDULLAHI2025116631, SI201716611, LARANJEIRA2025139, MANE202328076}, suggesting practical applicability.  

The desorption temperatures ($T_D$) ranged from 163.40 to 184.62 K, aligning with the weak adsorption strengths. Although these low $T_D$ values facilitate hydrogen release under mild conditions, they also imply a trade-off between easy desorption and ambient temperature stability. In particular, the system achieved a maximum HSC of 8.82 wt\% at 18H$_2$ saturation, surpassing the DOE target (5.5 wt\%) for viable hydrogen storage materials.  
 
\begin{table*}[h!]
\centering
\caption{Adsorption energy (E$_\mathrm{ads}$), average and maximum H--H bond length (R$_\mathrm{H-H}$ and R$_\mathrm{H-H_{Max}}$), desorption temperature (T$_D$), and hydrogen storage capacity (HSC) for K@GPD with different numbers of H$_2$ molecules.}
\label{tab:h2_adsorption}
\begin{tabular}{lccccc}
\hline
\textbf{System} & E$_\mathrm{ads}$ (eV) & R$_\mathrm{H-H}$ (\AA) & R$_\mathrm{H-H_{Max}}$ (\AA) & T$_D$ (K) & HSC (wt\%) \\
\hline
K@GPD--2H$_2$  & --0.13 & 0.77 & 0.77 & 163.40  & 1.06 \\
K@GPD--4H$_2$  & --0.11 & 0.76 & 0.77 & 141.08  & 2.11 \\
K@GPD--6H$_2$  & --0.13 & 0.76 & 0.77 & 162.70  & 3.13 \\
K@GPD--8H$_2$  & --0.12 & 0.76 & 0.77 & 151.98  & 4.12 \\
K@GPD--10H$_2$  & --0.12 & 0.77 & 0.83 & 153.62  & 5.10 \\
K@GPD--12H$_2$  & --0.14 & 0.77 & 0.86 & 176.87  & 6.06 \\
K@GPD--14H$_2$  & --0.14 & 0.77 & 0.82 & 184.62  & 7.00 \\
K@GPD--16H$_2$  & --0.13 & 0.76 & 0.80 & 169.78  & 7.92 \\
K@GPD--18H$_2$  & --0.14 & 0.76 & 0.79 & 180.68  & 8.82 \\
\hline
\end{tabular}
\end{table*}

The K@GPD--18H$_2$ configuration retains its metallic character, as evidenced by two bands crossing the Fermi level ($E_\mathrm{F}$) and increased dispersion in the occupied valence bands near $E_\mathrm{F}$. The PDOS analysis reveals substantial electronic reorganization: while C($p_z$) orbitals dominate near the Fermi level, notable contributions from H($s$), K($s$), and K($p_z$) orbitals emerge in the conduction region. This redistribution suggests a dual charge transfer mechanism in which K adatoms donate electrons to H$_2$ $\sigma^*$ antibonding states, facilitated by hybridization with C($p_z$) orbitals consistent with Kubas-type interactions.

Such interactions weaken the H–H bonds without dissociation, leading to moderate bond elongation. In this system, the average H–H bond lengths (R$_\mathrm{H-H}$) range from 0.76 to 0.77~\AA, with maximum values (R$_\mathrm{H-H_{Max}}$) reaching up to 0.86~\AA\ at higher coverages (see Table~\ref{tab:h2_adsorption}). These values significantly exceed the 0.75~\AA\ bond length of an isolated H$_2$ molecule, indicating a perturbation characteristic of Kubas-type binding~\cite{KAEWMARAYA2023157391}. The charge redistribution remains localized around the adsorbed H$_2$ molecules, preserving the overall metallicity of the system even at maximum hydrogen loading.

\begin{figure*}[pos=htb!]
    \centering
    \includegraphics[width=0.8\linewidth]{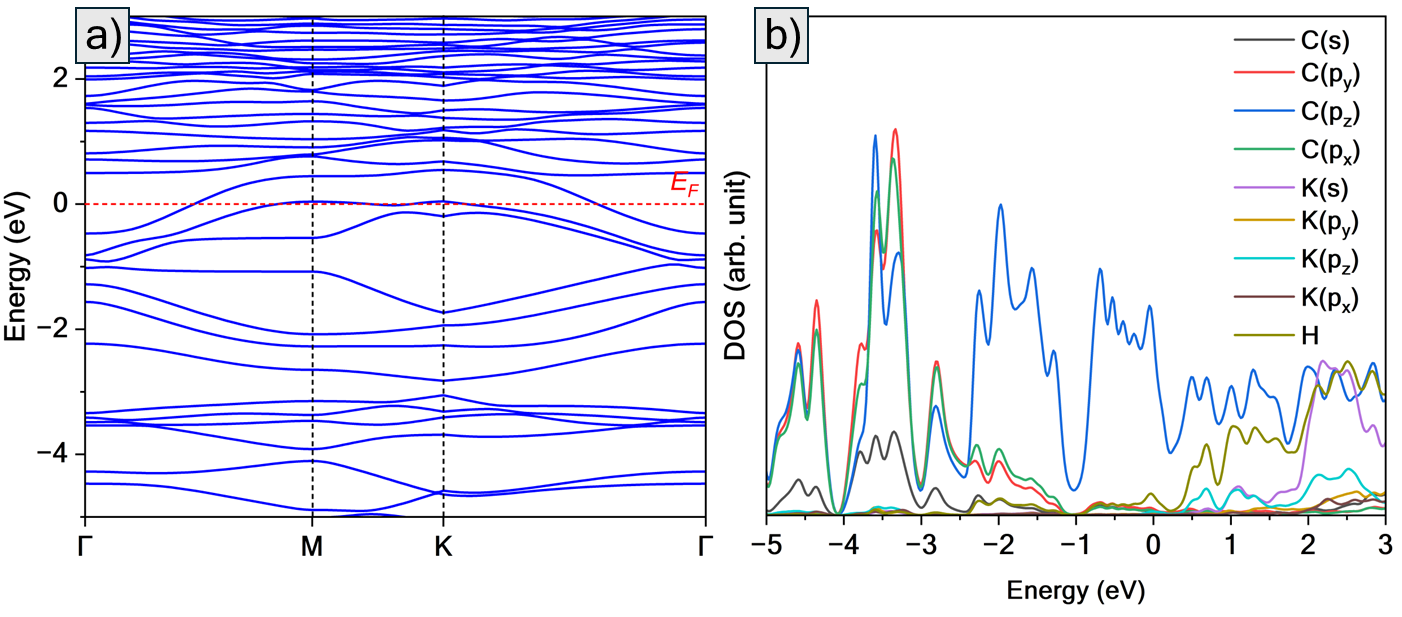}
    \caption{(a) Band structure and (b) PDOS for K@GPD--18H$_2$. The system remains the metallicity shown in the K@GPD complex, characterized by bands crossing the Fermi level ($E_F$ red dashed line).}
    \label{fig:enter-label}
\end{figure*}

\begin{figure}[pos=htb!]
    \centering
    \includegraphics[width=0.8\linewidth]{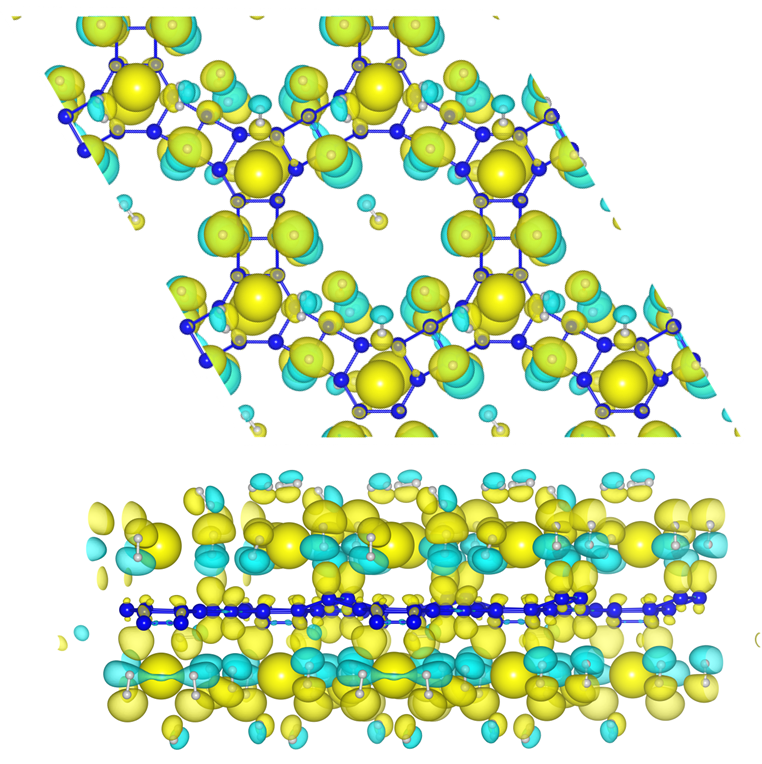}
    \caption{Top and side views of the CDD map for K@GPD--18H$_2$. The yellow (blue) regions indicate charge accumulation (depletion).}
    \label{fig:cdd2}
\end{figure}

Further insight into the charge redistribution mechanism is provided by the charge density difference (CDD) map for the K@GPD--18H$_2$ configuration, shown in Fig.~\ref{fig:cdd2}. The most prominent feature is the clear polarization of the H$_2$ molecules, characterized by localized regions of charge accumulation and depletion around the H–H bonds. This pattern supports the physisorptive nature of the H$_2$–K@GPD interaction, in agreement with the previously reported adsorption energies ranging from $-0.11$ to $-0.14$~eV. Complementary Bader charge analysis reveals a small yet significant charge transfer of approximately $-0.08$~$|e|$ per H$_2$ molecule from the potassium adatoms, corroborating the K($s$/$p_z$)–H($s$) hybridization observed in the PDOS.

\begin{figure*}[pos=htb!]
    \centering
    \includegraphics[width=0.8\linewidth]{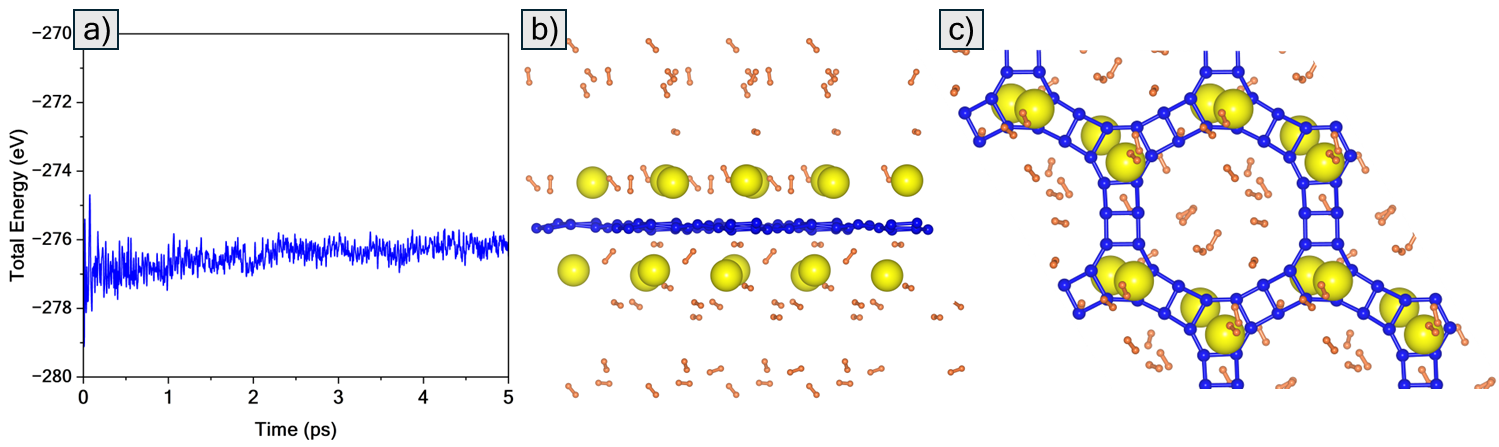}
    \caption{AIMD simulation results for K@GPD–18H$_2$ system at 300 K. (a) Time evolution of the potential energy and (b and c) final system configuration for K@GPD–18H$_2$. Irregular energy fluctuations indicate H$_2$ desorption events, demonstrating the reversible storage capability of K@GPD–18H$_2$.}
    \label{fig:aimd}
\end{figure*}

To evaluate thermal stability and desorption behavior, ab initio molecular dynamics simulations were conducted at 300~K for 5~ps (Fig.~\ref{fig:aimd}). The total energy profile shows dynamic fluctuations indicative of sequential H$_2$ desorption events, with several molecules detaching by the end of the simulation. This response confirms the system's reversible nature of hydrogen storage and aligns with the moderate adsorption strengths and low desorption temperatures (163–285~K) identified in static calculations. 

Crucially, the K@GPD substrate maintains its structural integrity throughout the simulation, with potassium adatoms retaining their original adsorption sites despite H$_2$ release. These findings highlight the reusability of K@GPD substrate, with K adatoms maintained on it regardless of multiple H$_2$ adsorption/desorption cycles without degradation.  

\begin{figure}[pos=htb!]
    \centering
    \includegraphics[width=0.8\linewidth]{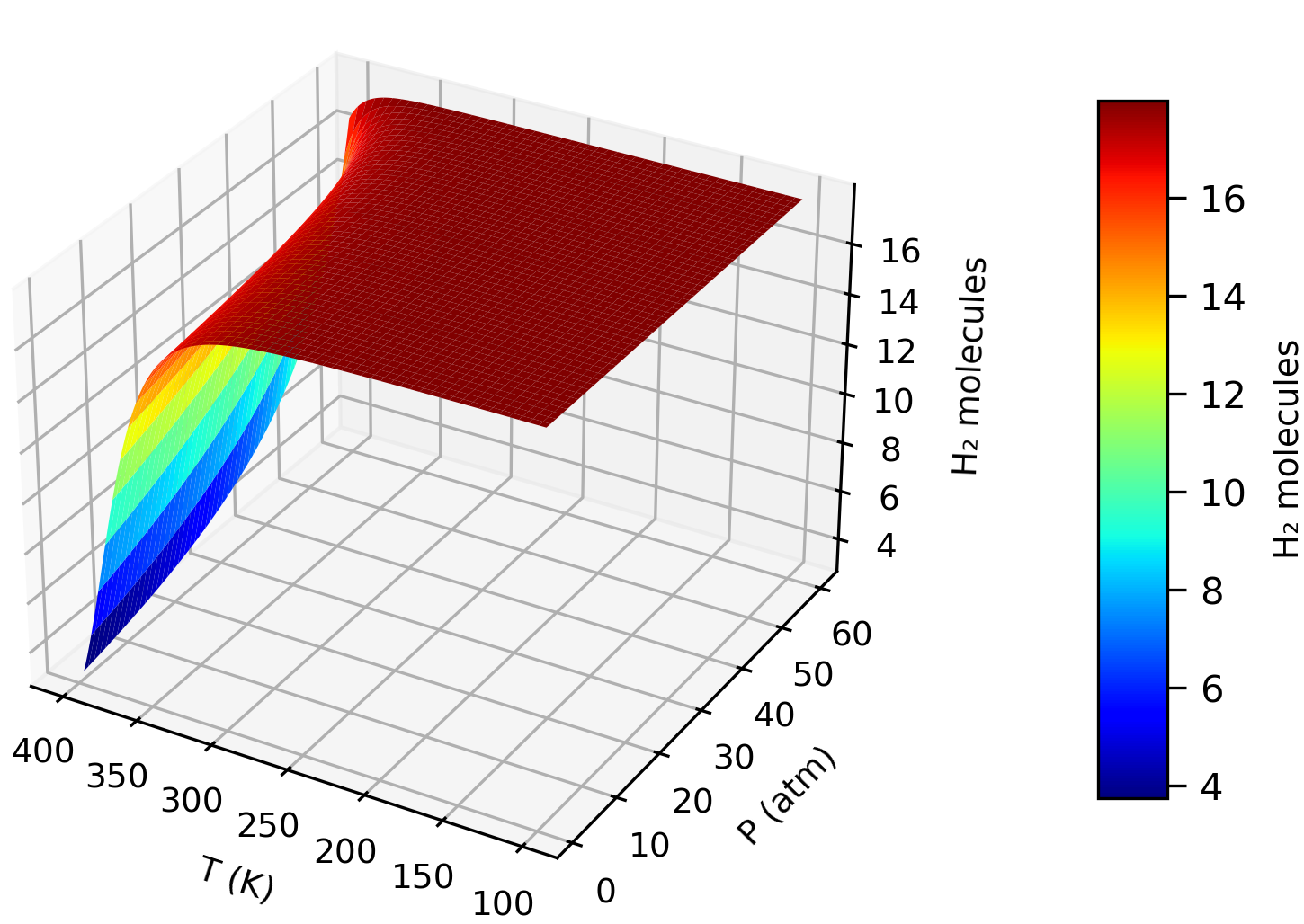}
    \caption{The average number of adsorbed H$_2$ on K@GPD at various temperatures (T) and pressures (P).}
    \label{fig:PTn}
\end{figure}

Figure~\ref{fig:PTn} presents the thermodynamic profile of hydrogen adsorption on K@GPD as a function of temperature and pressure. The analysis was conducted under representative operating scenarios: hydrogen uptake at 25~$^\circ$C and 30~atm, and release at 100~$^\circ$C and 3~atm. The practical storage capacity was determined by computing the difference in the number of H$_2$ molecules adsorbed under these conditions. Results show that 17.93 H$_2$ molecules are adsorbed at low temperature and high pressure, while 6.31 remain bound under desorption conditions, yielding a usable capacity of 11.62 molecules. This value corresponds to a gravimetric hydrogen storage capacity of 5.88~wt\%, exceeding the U.S. Department of Energy target of 5.5~wt\%. These findings highlight the potential of K@GPD as a viable material for reversible hydrogen storage applications.

Among recently proposed materials for hydrogen storage via metal decoration (Table~\ref{table:comparison}), the K@GPD monolayer exhibits a well-balanced performance. Although systems such as K@Aluminene achieve higher gravimetric capacities (9.41~wt\%), their extremely low adsorption energies (e.g., $-0.03$~eV) indicate weak physisorption and limited thermal stability. In contrast, K@GPD reaches 8.82~wt\%, surpassing well-known candidates including Li@$\alpha$-C$_3$N$_2$ (5.7~wt\%), Na@B$_7$N$_5$ (7.70~wt\%), and K@DHP-graphene (6.72~wt\%), while maintaining a moderate adsorption energy of 0.14~eV, within the optimal range for reversible storage under ambient conditions. Additionally, K@GPD offers one of the lowest desorption temperatures (181~K) among systems with comparable capacity. For instance, K@PHE-graphene, despite storing 7.47~wt\%, K@PHE-graphene requires desorption temperatures as high as 423~K.

\begin{table*}[!ht]
\label{table:comparison}
\centering
\caption{Total number of adsorbed H$_2$ molecules ($n$), Absolute adsorption energy per H$_2$ (|$\text{E}_\text{ads}$|), Hydrogen Storage Capacity (HSC), and Desorption temperature (T$_\text{des}$) associated with configurations exhibiting complete H$_2$ coverage configurations in recently documented systems.} 
\begin{tabular*}{\linewidth}{@{\extracolsep{\fill}} lcccc @{}}
\toprule
\textbf{System}                      & \textbf{n} & \textbf{|$\text{E}_\text{ads}$| (eV)} & \textbf{HSC (wt\%)} & \textbf{T$_\text{des}$ (K)} \\
\midrule
\textbf{K@GPD} (this work) & 18 & 0.14  & 8.82  & 181  \\
\textbf{Li@$\alpha$-C$_3$N$_2$} \cite{CHEN2024510} & 12 & 0.215  & 5.7  & 277 \\
\textbf{Na@B$_7$N$_5$} \cite{LIU2025105802} & 32  & 0.20  & 7.70  & 257  \\ 
\textbf{Na@Irida-graphene} \cite{duan2024reversible} & 32 & 0.14 & 7.82 & 195 \\
\textbf{K@BP-Biphenylene} \cite{djebablia2024metal} & 32 & 0.14 & 8.27 & - \\
\textbf{K@PHE-graphene} \cite{LARANJEIRA2025139} & 32 & 0.33 & 7.47 & 423 \\
\textbf{K@DHP-graphene} \cite{WANG2025141932} & 8 & 0.17 & 6.72 & - \\
\textbf{K@C$_9$N$_4$} \cite{KAUR202326301} & 7 & 0.17 & 8.1 & - \\
\textbf{K@Aluminene} \cite{VILLAGRACIA202116676} & 49 & -0.03 & 9.41 & - \\
\bottomrule
\end{tabular*}
\end{table*}

\section{Conclusions}

In this study, we employed DFT calculations to investigate the potential of potassium-decorated graphenyldiene (K@GPD) as a two-dimensional platform for reversible hydrogen storage. The pristine GPD monolayer, belonging to the $\mathrm{P6/mmm}$ space group, exhibits a direct band gap of 0.78~eV and a cohesive energy of $-6.92$~eV/atom, consistent with previous reports. Upon K-decoration, the system becomes metallic, with potassium atoms showing strong binding at energetically favorable adsorption sites, without any signs of clustering tendencies.

AIMD simulations confirmed the thermal stability of the K@GPD system near room temperature, with potassium adatoms remaining bound to the surface throughout the simulation. The hydrogen adsorption process was systematically analyzed by adding H$_2$ molecules incrementally. Adsorption energies ($E_\mathrm{ads}$) ranged from $-0.11$ to $-0.14$~eV, within the ideal range for reversible hydrogen storage and indicating physisorption behavior with minimal intermolecular interactions. The desorption temperatures ($T_D$) between 163 and 185~K suggest facile hydrogen release under mild conditions.

A maximum hydrogen storage capacity of 8.82~wt\% was achieved with 18 H$_2$ molecules, exceeding the U.S. DOE target of 5.5~wt\%. Electronic structure analysis revealed that H$_2$ adsorption induces significant charge redistribution and orbital hybridization, particularly involving C($p_z$), K($s$,$p_z$), and H($s$) states. The elongation of H--H bond lengths up to 0.86~\AA\ compared to the isolated value (0.75~\AA) provides strong evidence of Kubas-type interactions, wherein electron donation from K atoms to H$_2$ $\sigma^*$ orbitals facilitates adsorption without dissociation.

Finally, AIMD simulations of the saturated K@GPD--18H$_2$ system at 300~K demonstrated sequential H$_2$ desorption and confirmed the structural robustness and reusability of the K@GPD substrate. The resulting net usable capacity of 5.88~wt\% further highlights the promise of K@GPD as an efficient, lightweight, and regenerable material for future hydrogen storage applications.

\section*{Data access statement}
Data supporting the results can be accessed by contacting the corresponding author.

\section*{Conflicts of interest}
The authors declare no conflict of interest.

\section*{Acknowledgements}

 This work was supported by the Brazilian funding agencies Fundação de Amparo à Pesquisa do Estado de São Paulo - FAPESP (grant no. 22/03959-6, 22/16509-9, and 20/01144-0), National Council for Scientific and Technological Development - CNPq (grant no. 307213/2021–8 and 150187/2023-8), Coordination for the Improvement of Higher Education Personnel (grant no. 88887.827928/2023-00). The computational facilities were supported by resources supplied by the Molecular Simulations Laboratory (São Paulo State University, Bauru, Brazil). X.C. was funded by the Research Program of Chongqing Municipal Education Commission (No. KJQN202201327 and No. KJQN202301339), and the Natural Science Foundation of Chongqing, China (CSTB2022NSCQ-MSX0621).

\printcredits
\bibliography{refs}

\end{document}